\documentclass[a4paper,11pt]{article}
\pdfoutput=1 

\usepackage{jcappub} 

\usepackage[T1]{fontenc} 
\usepackage[utf8]{inputenc}
\usepackage{amssymb}
\usepackage{physics}
\usepackage{graphicx}
\usepackage{hyperref}
\usepackage{slashed}
\usepackage{bbm}
\usepackage{subcaption}
\usepackage{pifont}
\usepackage{xspace}

\usepackage[section]{placeins}

\newcommand{\be}{\begin{equation}}
\newcommand{\ee}{\end{equation}}
\newcommand{\bea}{\begin{eqnarray}}
\newcommand{\eea}{\end{eqnarray}}

\newcommand{\Lag}{\mathcal{L}}
\newcommand\ie{\mbox{\textit{i.\,e.}}\xspace}
\newcommand\cf{\mbox{c.\,f.}\xspace}
\newcommand\eg{\mbox{e.\,g.}\xspace}
\newcommand\D{\textrm{d}}

\usepackage{color}

\def \del{\partial}
\def\bea{\begin{eqnarray}}
\def\eea{\end{eqnarray}}
\newcommand{\nn}{\nonumber}

\title{Shift-symmetric Horndeski 
gravity in the 
asymptotic-safety paradigm
}

\author[a]{Astrid Eichhorn,}
\author[a]{Rafael R. Lino dos Santos}
\author[a,b,c]{and Fabian Wagner}

 \affiliation[a]{CP3-Origins,  University  of  Southern  Denmark,  Campusvej  55,  DK-5230  Odense  M,  Denmark}
\affiliation[b]{Institute of Physics, University of Szczecin, Wielkopolska 15, 70-451 Szczecin, Poland}
\affiliation[c]{Dipartimento di Ingegneria Industriale, Universit\`a degli Studi di Salerno, Via Giovanni Paolo II, 132 I-84084 Fisciano (SA), Italy}

\emailAdd{eichhorn@cp3.sdu.dk}
\emailAdd{rado@cp3.sdu.dk}
\emailAdd{fwagner@unisa.it}

\abstract{Horndeski gravity is a popular contender for a phenomenological model of dynamical dark energy, and as such subject to observational constraints. In this work, we ask whether Horndeski gravity can be more than a phenomenological model and instead become a fundamental theory, which extends towards high energy scales and includes quantum effects. We find that within the asymptotic-safety paradigm, an ultraviolet completion of a simple class of models of Horndeski gravity is achievable, but places strong constraints on the couplings of the theory. These constraints are not compatible with dynamical dark energy. Further, we find a similar result in an effective-field theory approach to this class of models of Horndeski gravity: under the assumption that there is no new strongly-coupled physics below the Planck scale, quantum gravity fluctuations force the Horndeski couplings to be too small to achieve an explanation of dynamical dark energy.}

\begin{document}
\maketitle
\flushbottom

\section{Introduction}
\label{sec:intro}

Several observations are challenging the CDM model, starting with the accelerated expansion of the universe \cite{SupernovaSearchTeam:1998fmf,SupernovaCosmologyProject:1998vns}, powered by an unknown dark energy. At the moment, this fact can be consistently accommodated by the addition of a cosmological constant. More recently, the $\sigma_8$ and Hubble tensions, the latter of which has grown to more than $4\sigma$ \cite{Planck:2018vyg,Riess_2021} have even been referred to as a "crisis in cosmology" \cite{DiValentino:2019qzk}.

To solve these tensions, and to replace the cosmological constant with a dark energy with dynamical origin, additional degrees of freedom are postulated. A popular class of these models consists of scalar-tensor-theories of gravity which add an independent scalar degree of freedom to the graviton (for a review, see \cite{Kobayashi:2019hrl}).

A straightforward way to avoid Ostrogradski instabilities, associated with a Hamiltonian operator that is unbounded from below, is to limit the equations of motion to at most second-order in time derivatives.\footnote{There are other, more subtle ways of avoiding instabilities. In the case of scalar-tensor-theories, this leads to the class of beyond-Horndeski models \cite{Gleyzes:2014dya,Gleyzes:2014qga} and degenerate higher-order scalar tensor theories \cite{Langlois:2015cwa}.} This constrains the possible terms that a healthy, classical, local, Lorentz covariant scalar-tensor Lagrangian can contain to those of Horndeski models \cite{Horndeski:1974wa,Deffayet:2009mn,Kobayashi:2019hrl}.

Horndeski theories are typically explored in the effective field theory framework \cite{Battye:2012eu, Baker:2012zs,Gubitosi:2012hu,Bloomfield:2012ff,Bloomfield:2013efa,Gleyzes:2013ooa,Koyama:2013paa,Gleyzes:2014rba,Langlois:2017mxy,Cusin:2017mzw, Solomon:2017nlh, Langlois:2018dxi}. Instead, we focus on their ultraviolet (UV) structure as a quantum theory. Our goal is twofold: first, we investigate whether Horndeski theories can be UV complete quantum field theories, in order to place them on a more solid theoretical footing. Second, we test whether the UV completion restricts the couplings in a Horndeski theory, in order to understand how the UV properties relate to phenomenologically interesting regions of the coupling space.\\

To reach this goal, we work in the asymptotic-safety paradigm, in which quantum field theories which satisfy appropriate conditions, can be UV complete and have predictive power.\footnote{It is an open question whether an asymptotically safe version of Horndeski gravity preserves the absence of instabilities. Within the truncation of the dynamics in which we work, such potential instabilities do not play a role.}

As this is the first such study in the context of asymptotically safe quantum field theories, we restrict ourselves to a technically simpler model, focusing on specific Horndeski theories, the kinetic braiding models, which describe dynamical dark energy \cite{Deffayet:2010qz,Kobayashi_2010,Pujolas:2011he,Creminelli:2017sry}. Additionally, we impose shift-symmetry of the scalar field $\phi,$ \ie, invariance under the transformation $\phi(x)\rightarrow\phi(x)+ a$, as investigated in \cite{Peirone:2019aua,Frusciante:2019puu,Traykova:2021hbr}. Shift symmetric models are special among asymptotically safe gravity-matter models, because their interactions cannot be set to zero \cite{Eichhorn:2012va,Eichhorn:2017eht,Laporte:2021kyp,deBrito:2021pyi}.

We test, using a truncation of the dynamics, whether the model belongs to the asymptotically safe landscape, \ie, the space of low-energy effective theories compatible with an asymptotically safe UV completion. Due to the high predictive power of asymptotic safety, this landscape is expected to be quite small and only contain a limited number of theories; restricted not only in their matter content \cite{Dona:2013qba}, but also in their allowed interactions \cite{Eichhorn:2017eht, Eichhorn:2017lry,Eichhorn:2018whv,Eichhorn:2020sbo,deBrito:2021akp}. More general swampland-conjectures from quantum gravity have also been investigated in asymptotic safety, see \cite{deAlwis:2019aud,Basile:2021krr}.  For introductions and reviews on asymptotic safety for gravity and matter, see \cite{Eichhorn:2018yfc,
Eichhorn:2022jqj, Eichhorn:2022gku}. For discussions of asymptotic safety in cosmology, see, e.g., \cite{Bonanno:2017pkg,Platania:2020lqb,Wetterich:2022ncl} for reviews.
\\ 
 
The paper is structured as follows: we introduce the concept and formalism of asymptotic safety, focusing on gravity-matter models in Sec.~\ref{sec:ASQG}. We motivate our truncations of the full dynamics and connect the couplings to observable parameters in Sec.~\ref{sec:Model}. We analyze the results from asymptotic safety and effective field theory in Sec.~\ref{sec:resultsAS}, and finally close with a discussion and an outlook in Sec.~\ref{sec:results}. We provide further technical details and study the effect of a higher-order operator in the appendix.

\section{Asymptotically safe quantum gravity}
\label{sec:ASQG}

Historically, perturbative renormalizability has been the main tenet for the choice of valid quantum field theories.
However, it was already recognized early on that perturbative renormalizability is neither necessary nor sufficient for a theory to be UV complete. Theories which are perturbatively renormalizable can nevertheless feature Landau poles and therefore be trivial. Conversely, theories which are not perturbatively renormalizable may nevertheless be UV complete by being asymptotically safe. This is an option both in and beyond perturbation theory. For this to occur, the running of the couplings is determined by a competition of screening and antiscreening quantum fluctuations. These balance out at an interacting (or non-Gaussian) fixed point. The theory therefore exhibits quantum scale symmetry in the far UV. The scale-symmetric regime connects to IR physics (which typically contains physical scales) by Renormalization Group (RG) trajectories that depart from the fixed point.
This behavior has been observed for instance in non-linear sigma and Gross-Neveu models in $d=2+\epsilon$ \cite{Brezin75,Gawedzki85}, in Yang-Mills theories in $d=4+\epsilon$ \cite{Peskin:1980ay},  and in weakly-coupled gauge-Yukawa models in $d=4$ \cite{Litim:2014uca} (see \cite{Eichhorn:2018yfc} for a review and references). 

In \cite{Weinberg79} the asymptotic-safety paradigm was suggested for gravity, because it could provide a predictive quantum field theory despite gravity's perturbative nonrenormalizability. 
According to the paradigm, an infinite number of couplings -- required to describe gravity in the UV -- would become interdependent, leading to a finite number of free parameters and, thus, a predictive theory. Starting from the seminal paper
 \cite{Reuter:1996cp} which adapted functional RG (FRG) techniques to gravity, an interacting fixed point was discovered in \cite{Souma:1999at}. This finding has been corroborated in increasingly less severe approximations to the dynamics of the theory \cite{
Lauscher:2001ya,Reuter:2001ag,Lauscher:2002sq,Litim:2003vp,Codello:2006in,Machado:2007ea,Codello:2008vh,Benedetti:2009rx,Eichhorn:2009ah,Manrique:2010am,Eichhorn:2010tb,Groh:2010ta,Dietz:2012ic,Christiansen:2012rx,Rechenberger:2012pm,Falls:2013bv,Ohta:2013uca,Eichhorn:2013xr,Falls:2014tra,Codello:2013fpa,Christiansen:2014raa,Demmel:2015oqa,Gies:2015tca,Christiansen:2015rva,Ohta:2015fcu,Ohta:2015efa,Falls:2015qga,Eichhorn:2015bna,Gies:2016con,Denz:2016qks,Biemans:2016rvp,Falls:2016msz,Falls:2016wsa,deAlwis:2017ysy,Christiansen:2017bsy,Falls:2017lst,Houthoff:2017oam,Falls:2017cze,Becker:2017tcx,Knorr:2017fus,Knorr:2017mhu,DeBrito:2018hur,Eichhorn:2018ydy,Falls:2018ylp,Bosma:2019aiu,Knorr:2019atm,Falls:2020qhj,Kluth:2020bdv,Bonanno:2021squ,Knorr:2021slg,Baldazzi:2021orb,Sen:2021ffc,Mitchell:2021qjr,Knorr:2021iwv,Fehre:2021eob,Baldazzi:2021fye,Sen:2022xlp}, see \cite{Eichhorn:2017egq,Percacci:2017fkn, Reuter:2019byg,Pereira:2019dbn,Reichert:2020mja,Pawlowski:2020qer} for reviews. The open challenges of this program have been discussed in \cite{Donoghue:2019clr,Bonanno:2020bil} and recent progress on two key questions, unitarity, and Lorentzian signature, has been achieved in, e.g., \cite{Draper:2020bop,Platania:2020knd,Fehre:2021eob}. 

\subsection{The predictive power of asymptotic safety}
It is a key strength of the asymptotic-safety paradigm that it comes with predictive power. There are even indications that free parameters of the Standard Model are calculable from first principles by demanding asymptotic safety of the Standard Model with gravity \cite{Shaposhnikov:2009pv,Harst:2011zx,Eichhorn:2017ylw,Eichhorn:2017lry,Eichhorn:2018whv,Alkofer:2020vtb,Kowalska:2022ypk,Eichhorn:2022vgp}.

At an intuitive level, the predictive power comes from an extra symmetry, namely quantum scale symmetry in the UV. Similarly to other symmetries, it requires relations between couplings and thus enhances the predictive power over a model which does not feature the symmetry. Differently from classical symmetries, quantum scale symmetry only holds in the UV, not in the IR, but its predictive power extends over all scales, as we will explain below.

At the technical level, the predictive power comes from the scaling properties of the couplings: only relevant couplings can run away from scale symmetry under the RG flow to the IR. These relevant couplings are distinguished by positive critical exponents, as we review now.

The beta function of a coupling $\bar{g}_{i}$ is given by the logarithmic scale-derivative of the coupling, $\beta_{\bar{g}_i} = k \partial_k\, \bar{g}_i$. To discover asymptotic safety or scale-symmetry, it is most convenient to remove all explicit scales from the theory by working with dimensionless couplings $g_i = k^{-d_{\bar{g}_i}}\bar{g}_i$, where $d_{\bar{g}_i}$ is the canonical mass-dimension of a coupling.
A fixed point, indicated by the subscript $*$ in the couplings, is defined as a real root of the beta function
\begin{equation}
\left.\beta_{g_i}\right|_{g_{i\,\ast}}\equiv 0.
\end{equation}
 All quantum field theories feature a Gaussian, \ie, noninteracting, fixed point, because quantum fluctuations cannot have a screening or antiscreening effect if all interactions vanish. 
 In contrast, interacting fixed points are not guaranteed in a generic quantum field theory, but are known in many examples.

Analyzing the behavior of couplings in the immediate neighbourhood of a fixed point determines the relevant couplings. To that end, one expands the beta function to leading order in the distance to the fixed point, $g_i-g_{i\, \ast}$,
\begin{equation}
\beta_{g_i}=\beta_{g_i}|_{g_{i\,\ast}} + \sum_j M_{ij}\left(g_j-g_{j\, \ast}\right) + \mathcal{O}\left(g_i-g_{i\,\ast} \right)^2,\label{eq:linearizedbeta}
\end{equation}
with the stability matrix $M_{ij}\equiv\partial\beta_{g_i}/\partial g_j|_{g_i^*}.$ 
We define the eigenvalues of $M$, multiplied by an additional negative sign, as the critical exponents, 
\be
\theta_I = - {\rm eig}M.
\ee
They enter the solution of Eq.~\eqref{eq:linearizedbeta}, which is
\begin{equation}
g_i=g_{i\,\ast}+\sum_J c_J V_i^J\left(\frac{k}{k_0}\right)^{-\theta_J},
\end{equation}
where the $V^J$ are the eigenvectors of the stability matrix and the $c_J$ are integration constants, while $k_0$ stands for an arbitrary  reference scale.

To explain the effect of a positive critical exponent, we assume for simplicity that the eigenvalues $V^J$ are aligned with the couplings (otherwise the same argument holds for superpositions of the original couplings $g_i$.)
Under the RG flow from the UV to the IR, i.e., from high to low $k$, a positive critical exponent makes a deviation from scale-symmetry possible: If $c_J$ is chosen to be non-zero at some reference scale $k_0$, $g_i(k<k_0)$ deviates more and more from $g_{i\, \ast}$. The constant $c_J$ therefore determines the low-energy value of $g_i$; thus, each positive critical exponent translates into one free parameter of the theory.

In contrast, a negative critical exponent makes a deviation from scale-symmetry impossible: irrespective of the value of $c_J$, $g_i(k<k_0)$ is driven towards $g_{i\, \ast}$. The low-energy value of $g_i$ is independent of $c_J$; thus each negative critical exponent removes one of the free parameters that an effective field theory without a fixed point would have.

At non-interacting fixed points, the critical exponents are equal to the mass dimensions of the couplings, such that there is only a finite number of relevant directions. At interacting fixed points, the critical exponents depend on the mass dimensions, but also on finite contributions from quantum fluctuations. The scaling spectrum, i.e., the set of critical exponents, is therefore shifted (not necessarily by the same amount in each critical exponent) in comparison to the set of mass dimensions. There is evidence that this shift is relatively small for an asymptotically safe gravity-matter fixed point \cite{Falls:2013bv,Falls:2014tra,Falls:2018ylp,Eichhorn:2020sbo} and that it is negative for at least some matter couplings, see, e.g., \cite{Harst:2011zx,Wetterich:2016uxm,Eichhorn:2017lry,Eichhorn:2017als,Eichhorn:2017ylw,Eichhorn:2018whv,Pawlowski:2018ixd,Eichhorn:2020sbo,Eichhorn:2021tsx}, such that the predictivity of an asymptotically safe Standard Model may be increased compared to the Standard Model without gravity and asymptotic safety.

\subsection{Interplay of asymptotically safe gravity and matter}\label{sec:interplay}

The observed matter fields must be included in a quantum-gravity theory consistent with our universe. This poses a challenge for some approaches to quantum gravity. Therefore, it is a nontrivial test of a quantum-gravity model, whether or not it can accommodate the observed matter fields (without predicting additional fields that should have already been observed).
In asymptotically safe quantum gravity, it is not a priori guaranteed that the fixed-point structure of the pure-gravity sector is preserved when including matter. Non-trivial evidence is mounting for asymptotic safety in gravity-matter models with Standard-Model fields and interactions, e.g., \cite{
Dona:2013qba,Dona:2014pla,Meibohm:2015twa,Oda:2015sma,
Wetterich:2016uxm,
Biemans:2017zca,
Eichhorn:2017eht,Eichhorn:2017ylw,Christiansen:2017cxa,Hamada:2017rvn,
Eichhorn:2018ydy,Alkofer:2018fxj,
Wetterich:2019zdo,
deBrito:2019umw,Kwapisz:2019wrl,Eichhorn:2020sbo,
Sen:2021ffc, deBrito:2022vbr,Pastor-Gutierrez:2022nki}. Additional studies focusing on scalar-gravity models \cite{Narain:2009fy,Eichhorn:2012va,Dona:2015tnf,Percacci:2015wwa, Labus:2015ska,Eichhorn:2017sok,Pawlowski:2018ixd,Eichhorn:2018akn,Burger:2019upn,deBrito:2021pyi,Laporte:2021kyp,Knorr:2022ilz}, fermion-gravity models \cite{Eichhorn:2011pc,Dona:2012am,Eichhorn:2016vvy,Eichhorn:2018nda,deBrito:2019epw,deBrito:2020dta,Daas:2021abx} and vector-gravity models \cite{Daum:2009dn,Harst:2011zx,Folkerts:2011jz,Eichhorn:2017lry,Eichhorn:2019yzm,Eichhorn:2021qet,Wetterich:2022bha} provide additional support for asymptotic safety in gravity-matter models. Open questions on gravity-matter models are reviewed together with the state-of-the-art in \cite{Eichhorn:2022gku}.
Additionally, lattice techniques have been used to investigate the interplay of gravity and matter with a view towards asymptotic safety \cite{Jha:2018xjh,Catterall:2018dns,Dai:2021fqb,Ambjorn:2021fkp,Ambjorn:2021uge}.

A key outcome of such studies is a concrete suggestion for observational tests of quantum gravity. These rely on negative critical exponents for Standard-Model like interactions. Once UV scale symmetry is required, each negative critical exponent imposes one additional constraint on the dynamics of the theory. Concretely, this constraint fixes one initial condition for the RG flow in the Standard Model at the Planck scale. Once this initial condition is fixed, the  value at low energies (e.g., at the electroweak scale), which is accessible to measurements, is determined. This is in contrast to the Standard Model without quantum gravity and asymptotic safety, where the Planck-scale values of the couplings are free parameters. Indications for this mechanism exist \cite{Harst:2011zx,Eichhorn:2017lry,Eichhorn:2017ylw,Eichhorn:2017als,Pawlowski:2018ixd,Wetterich:2019rsn,
Eichhorn:2020sbo}, giving rise to an asymptotically safe phenomenology in (and beyond) the Standard Model \cite{Shaposhnikov:2009pv,Eichhorn:2018whv,Grabowski:2018fjj,Eichhorn:2019dhg,Reichert:2019car,Alkofer:2020vtb,
Hamada:2020vnf,Eichhorn:2020kca,Kowalska:2020gie,Domenech:2020yjf,Kowalska:2020zve,Eichhorn:2021tsx,deBrito:2021akp,Kowalska:2022ypk,Eichhorn:2022vgp,Held:2022hnw,Wetterich:2022bha,Boos:2022jvc,Pastor-Gutierrez:2022nki,Chikkaballi:2022urc,Boos:2022pyq}.

In addition to the constraints that asymptotic safety imposes at low energies, there could also be a constraint that is active above the Planck scale.  In turn it could constrain the dimensionality of spacetime to four \cite{Eichhorn:2019yzm} and may also impose bounds on the number of matter fields \cite{deBrito:2021pyi,Laporte:2021kyp}. This constraint is called the weak-gravity bound (unrelated to the so-called weak-gravity conjecture) and limits the allowed values of gravitational couplings at the fixed point.
The gravitational couplings must be small enough at the fixed point for gravity to be sufficiently weakly coupled, otherwise gravitational fluctuations spoil the presence of fixed points in the matter sector \cite{Eichhorn:2012va, Eichhorn:2016esv,Christiansen:2017gtg,Eichhorn:2017eht,
Eichhorn:2019yzm, deBrito:2021pyi,Laporte:2021kyp,Eichhorn:2021qet,Knorr:2022ilz}, at least in those truncations of the dynamics studied to date. 

For both sets of constraints, those related to negative critical exponents and those related to the weak-gravity bound, several caveats hold \cite{Bonanno:2020bil}. First, these results are obtained  in systematic approximations of the RG flow which are expected to capture the leading gravitational contributions. However, (apparent) convergence of results under subsequent refinements of the approximation has not been demonstrated yet. Second, these results are obtained in Euclidean spacetime signature which in quantum gravity is not straightforwardly related to the phenomenologically relevant Lorentzian signature \cite{Manrique:2011jc,Draper:2020bop,Platania:2020knd,Bonanno:2021squ,Fehre:2021eob,Knorr:2022mvn}.
 
\subsection{The functional renormalization group}
Asymptotic safety in gravity is  explored with dynamical triangulations \cite{Loll:2019rdj,Ambjorn:2020rcn}, tensor-models \cite{Eichhorn:2018phj,Eichhorn:2019hsa} and in spin-foam settings \cite{Bahr:2016hwc,Steinhaus:2020lgb}. However, most progress has been achieved using functional RG techniques, see \cite{Eichhorn:2022jqj} for a recent review, which are therefore our method of choice.

In the functional RG, beta functions are extracted from a flow equation. This flow equation governs the scale dependence of the effective dynamics, encoded in a scale-dependent effective action, $\Gamma_k$. This scale-dependent effective action is the quantum counterpart of the classical action, and consists of field monomials multiplied by scale-dependent couplings. In contrast to the classical action, the effective action contains all interactions compatible with the symmetries, because these are all generated by quantum fluctuations. Therefore, the scale-dependent effective action also contains all possible interactions. This turns it into an ideal tool to calculate the scale dependence of all couplings in the theory (not just those related to perturbatively renormalizable interactions).

This flow equation \cite{Wetterich:1992yh,Morris:1993qb} reads
\begin{equation}
k\partial_k\Gamma_k=\frac{1}{2}\text{Tr}\left(
k\partial_k R_k
\left(\Gamma_k^{(2)}+R_k\right)^{-1}\right),\label{FRGeq}
\end{equation}
where the trace applies to all internal and Lorentz indices, as well as the spacetime/momentum dependence and the field content. The regulator function $R_k$ endows low-momentum quantum fluctuations with a mass-like term, suppressing their contribution to $\Gamma_k$. In contrast, high-momentum configurations are integrated out. 
The superscript $(2)$ denotes the second variational derivative of the scale-dependent effective action with respect to the fields. The equation is formally exact and applies both within perturbation theory and beyond. It has a one-loop structure, but is not limited to reproducing one-loop results, see, e.g., \cite{Papenbrock:1994kf}.

From an ansatz for the effective dynamics one can extract the beta functions. Such an ansatz takes the general form
\be
\Gamma_k = \sum_i \bar{g}_i \int_x \mathcal{O}_i(\Phi),
\ee
where $\int_x$ stands for the Euclidean spacetime integral, dimensionful couplings multiply the various field monomials $\mathcal{O}(\Phi)$, and $\Phi$ is a placeholder for the fields in the theory. The scale-derivative of $\Gamma_k$ therefore contains the beta functions for the dimensionful couplings,
\be
k\, \partial_k\, \Gamma_k = \sum_i \beta_{\bar{g}_i} \int_x \mathcal{O}_i(\Phi).
\ee
Therefore, by projecting onto a given field monomial, one can extract the corresponding beta function. The beta functions can be used in two ways: first, by searching for common roots of all beta functions, one can find fixed points. Second, by providing an initial condition (in the form of specifying values for the couplings at some high scale), one can use the beta functions to integrate from this high scale to low energies and thus connect a UV fixed point regime to an IR regime in which experimental measurements exist.

In contrast to approaches that only keep track of perturbatively renormalizable couplings, the FRG keeps track of all couplings. The underlying reason is that quantum fluctuations generate all terms compatible with the symmetries in the effective action. This is also well-known from perturbation theory, where the one-loop effective action contains higher-order terms. In a particle-physics context, such terms have recently come into focus in the context of the Standard-Model effective field theory \cite{Brivio:2017vri}. 

In practice, keeping track of all (infinitely many) couplings is not possible; thus, an ansatz for $\Gamma_k$ (also called a truncation) is chosen. The systematic error that is made by neglecting some of the couplings cannot always be quantified. Therefore, it is essential to observe recurring, \ie, robust patterns while the truncation is successively extended. Despite the need for an approximation (which, in one or another form is also necessary with other QFT techniques, be they perturbative or non-perturbative), the FRG is applied very successfully, and also with quantitative precision, in a broad range of areas from statistical physics through condensed matter and particle physics to quantum gravity, see \cite{Dupuis:2020fhh} for a recent review.

\section{A first Horndeski-like model in the asymptotic-safety paradigm  -- setup}
\label{sec:Model}

Horndeski theories of gravity fall under the general class of scalar-tensor theories. In particular, they complement the field content of general relativity with a scalar $\phi(x)$ in such a way that the resulting equations of motion are of second order in derivatives, thereby ameliorating the risk of dynamical instabilities  at the classical level.\footnote{We highlight that stability of a classical action does not imply that the corresponding quantum theory is ghost-free at an asymptotically safe fixed point. Additional interactions which are generated by quantum fluctuations might generate ghosts.} Under this assumption, the underlying Lagrangian (here expressed in four-dimensional Euclidean spacetime) has to be constructed from the operators
\begin{align}
\Lag_2= \,&-G_2(\phi,\chi),\\
\Lag_3=\, &G_3(\phi,\chi)\Box\phi,\\
\Lag_4=\, &-G_4(\phi,\chi)R+G_{4,\chi}\left[\left(\Box\phi\right)^2-\nabla_\mu\nabla_\nu\phi\nabla^\mu\nabla^\nu\phi\right],\label{L4}\\
\Lag_5=\, &G_5(\phi,\chi)G_{\mu\nu}\nabla^\mu\nabla^\nu\phi\nonumber\\
			\,&-\frac{G_{5,\chi}}{6}\left[\left(\Box\phi\right)^3-3\Box\phi\nabla_\mu\nabla_\nu\phi\nabla^\mu\nabla^\nu\phi+2\nabla_\mu\nabla_\nu\phi\nabla^\mu\nabla^\rho\phi\nabla_\rho\nabla^\nu\phi\right],
\end{align} 
where $G_2,G_3,G_4,G_5$ are functions of the kinetic term $\chi=-\nabla_\mu\phi\nabla^\mu\phi/2$ and the field $\phi$ itself. Commas indicate partial derivatives. Furthermore, the Lagrangian contains the Ricci scalar $R$ as well as the Einstein tensor $G_{\mu\nu}$. 

 Different choices for $G_2,G_3,G_4,G_5$ correspond to physically distinct theories of modified gravity and dark energy, see for instance \cite{Zumalacarregui:2012us}. 
In order to analyze the viability of a Horndeski model in the asymptotic-safety paradigm, we choose a particular truncation of the effective dynamics.   

\subsection{Truncation: shift-symmetric kinetic braidings}
\label{subsec:trunc}
To choose our truncation, we are guided by technical simplicity and by experimental bounds. The operators $\Lag_4$ and $\Lag_5$ change the dispersion relation of gravitational waves which has been constrained\footnote{This may be subject to caveats regarding the validity of the effective field theory \cite{deRham:2018red}.} by the first gravitational wave detection with electromagnetic counterpart GW170817 \cite{LIGO17a,Goldstein17,LIGO17b}. In particular, to satisfy the observational bounds that the speed of gravitational waves is very close to the speed of light, $G_{4,\chi}, G_5$ and $G_{5,\chi}$ must be very close to zero \cite{Creminelli:2017sry,Ezquiaga:2017ekz,Sakstein17,Baker17,Amendola17,Crisostomi17,Arai:2017hxj,Ishak:2018his,Kase18a,Copeland:2018yuh, Baker:2020apq}. Therefore, from now on, we focus on the class of kinetic braidings \cite{Deffayet:2010qz,Pujolas:2011he}, based on the assumptions  $G_{4,\chi}=0$ and $G_5=0$.

 One may argue that $G_{4,\chi}\approx 0$ and $G_5 \approx 0$ hold generically: both couplings are of mass dimension -2 (or higher). Thus, even if they are nonzero at high scales, they are expected to be suppressed at low scales. Accordingly, the observation that constrains these couplings to be small may be an automatic consequence of the Renormalization Group flow in such a theory. We confirm that such an expectation holds for some of the other interactions in Horndeski theories that we keep in our truncation.

We further impose shift symmetry, \ie, $\phi (x) \rightarrow \phi (x) + a$ for simplicity. Correspondingly, the scale-dependent effective action can only depend on the field through its derivatives, i.e., $G_2(\phi, \chi) \rightarrow G_2(\chi)$, $G_{3}(\phi, \chi) \rightarrow G_3(\chi)$ and $ G_4(\phi)	\rightarrow G_4\propto M_{Pl}^2$. In \cite{Narain:2009fy,Eichhorn:2012va,Eichhorn:2017als,deBrito:2021pyi,Laporte:2021kyp}, evidence has accumulated that scalar-field interactions which break shift symmetry are not generated by asymptotically safe quantum gravity and may thus be set to zero consistently. Shift-symmetric Horndeski models were previously considered in the context of cosmological \cite{Peirone:2019aua,Frusciante:2019puu,Traykova:2021hbr} and hairy black hole solutions \cite{Kobayashi:2014eva, Khoury:2020aya, Khoury:2022zor}. 

We limit the truncation to operators which are at most of mass dimension 8 and, therefore, work to second order in the polynomial expansion of $G_2$ and to first order in the expansion of $G_3$. The resulting model has been previously considered in \cite{Traykova:2021hbr}.  We further introduce the wave function renormalization $Z_\phi$. Thus, our truncation of the scale-dependent effective action is
\begin{equation}
\Gamma_{k,\text{Horndeski}}=\int\D^4x\sqrt{\det g_{\mu\nu}} \left[-\frac{1}{16\pi \bar{G}}\left(R-2\bar{\Lambda}\right)-Z_{\phi}\chi-\bar{h}
\chi\Box\phi+\bar{g}
\chi^2\right],\label{horndeskiacp}
\end{equation}
with the dimensionful scalar couplings $\bar{h}$ and $\bar{g},$ the determinant of the metric $\det(g_{\mu\nu})$, the Newton coupling $\bar{G}$ and the cosmological constant $\bar{\Lambda}$. While the latter is kept for reasons of generality, in the context of Horndeski theories, it should ideally be made redundant by the scalar at IR scales. In previous works on asymptotically safe gravity, $\bar{\Lambda}$ comes out as a relevant coupling \cite{Eichhorn:2018yfc}, thus its IR value can be chosen to vanish, consistent with a Horndeski-like phenomenology. Additionally, we introduce a gauge-fixing term for gravitational fluctuations, see App.~\ref{app:beta}.

Rewritten in terms of the dimensionless couplings which are also rescaled by an appropriate factor of the wave-function renormalization, Eq.~\eqref{horndeskiacp} becomes
\begin{align}
\Gamma_{k,\text{Horndeski}}	=&\int d^4 x \sqrt{\det g_{\mu\nu}}\left[-\frac{1}{16\pi G \,k^{-2}}\left(R-2\Lambda\,k^2\right)+\frac{Z_\phi}{2}g^{\mu\nu} \del_\mu\phi\del_\nu\phi \right. \nn \\
	& \left. +\frac{Z_\phi^{3/2} \,h\,k^{-3}}{2}g^{\mu\nu}g^{\alpha\beta}\del_\mu\phi\del_\nu\phi\nabla_\alpha\del_\beta\phi+\frac{Z_\phi^2\, g \,k^{-4}}{4}g^{\mu\nu}g^{\alpha\beta}\del_\mu\phi\del_\nu\phi\del_\alpha\phi\del_\beta\phi\right].\label{effac}
\end{align}
We note that due to the one-loop structure of the flow equation, $h$ and $g$ cannot enter the beta functions for $G$ and $\Lambda$ directly. Therefore, we only compute the beta functions for $h$ and $g$ in this work, extending \cite{deBrito:2021pyi} which contained only $g$, but not $h$.

\subsection{From couplings to observable parameters}\label{sec:obs}

Couplings in a QFT do typically not correspond to observables; instead, they enter observables in various combinations. This is important to keep in mind in the development of quantum-gravity phenomenology. It also matters in the interpretation of beta functions: these are not universal beyond two loops for canonically dimensionless couplings and not universal at all for canonically dimensionful couplings. Accordingly, the IR value of a coupling that results from an integrated RG flow is also not necessarily universal; thus, beta functions as well as values of couplings can only be compared across different schemes with due care.\\

To derive phenomenological constraints and review observational bounds on the couplings of the model, we work in Lorentzian signature in this subsection. We use the mostly-positive sign convention for the metric. We will later use our results obtained from RG flows in Euclidean signature and compare directly to the constraints discussed below. This is justified under the assumption that the results for the RG flow remain the same under a change of signature.\\

Below, we apply two sets of constraints. First, we consider bounds which arise by requiring that the scalar field provides dynamical dark energy. Kinetic braiding models can dynamically describe dark energy for an appropriate range of values of the couplings $\bar{h}$ and $\bar{g}$. Second, we use constraints on the Horndeski-parameters, which are combinations of the couplings that can be accessed by measurements. Together, these provide us with constraints on $\bar{h}$ and $\bar{g}$ which we compare to the theoretical constraints on the couplings that arise from an embedding of the model into the asymptotic-safety paradigm.

We start off with the functional
\begin{equation}
	S_{\text{Horndeski}}=\int\D^4x\sqrt{-\det g_{\mu\nu}} \left[\frac{R}{16\pi \bar{G}}+\chi+\bar{h}\chi\Box\phi-\bar{g}\chi^2\right],\label{LorAc}
\end{equation}
with $\chi=-\nabla_\mu\phi\nabla^\mu\phi/2$ now understood in Lorentzian signature.

In order for the scalar to replace the cosmological constant, its equation of state parameter  
$w$, which can be obtained from the action Eq.~\eqref{LorAc},
\begin{equation}
		w=-1+\frac{2+2\bar{h}(\ddot{\phi}-3H\dot{\phi})-2\bar{g}\dot{\phi}^2}{1-6\bar{h}H\dot{\phi}-\frac{3}{2}\bar{g}\dot{\phi}^2},
\end{equation}
has to satisfy $w_0\simeq -1.0$ \cite{Planck:2018vyg} at present times (indicated by the subscript zero).
Herein, $H$ is the Hubble parameter and time-derivatives are denoted by a dot. 
Thus
the couplings have to satisfy 
\begin{equation}
	\bar{h}[3\omega_0 H_0\dot{\phi}_0+\ddot{\phi}_0]+\frac{3\omega_0 -1}{4}\bar{g}\dot{\phi}_0^2= \frac{\omega_0 -1}{2},\label{wcond}
\end{equation}
with the Hubble constant $H_0\simeq 8.9\times 10^{-33}\text{eV}$  (see \cite{Planck:2018vyg}). Additionally, the energy density of the scalar field at present has to equal the measured dark energy density, 
\begin{equation}
		\frac{3H_0^2\Omega_{\Lambda,0}}{8\pi 
			\bar{G}
		}= \frac{\dot{\phi}_0^2}{2}\left(1-6\bar{h}H_0\dot{\phi}_0-\frac{3\bar{g}}{2}\dot{\phi}_0^2\right)\label{Friedcond},
\end{equation}
with the density parameter of dark energy at present $\Omega_{\Lambda,0}\simeq 0.69$ \cite{Planck:2018vyg}. 
 
We first set $\bar{h}=0$, which will become important in Sec.~\ref{sec:resultsAS}. Then, Eqs.~\eqref{wcond} and \eqref{Friedcond} imply that
\begin{equation}
		\frac{3H^2_0\, \Omega_{\Lambda,0}}{8\pi 
	\bar{G}
	} = -\frac{\dot{\phi}_0^2}{4}.\label{neccosmconst}
\end{equation}
The energy density on the right-hand side is negative which disagrees with observations \cite{Planck:2018vyg}. Thus, if $\bar{h}=0,$ the model cannot serve as a dynamical explanation of dark energy. This result will provide an important constraint on quantum gravity.

For $\bar{h}\neq 0$, the situation is more involved, because Eq.~\eqref{wcond} depends on $\ddot{\phi}$. Therefore, we also use the equations of motion, given by 
\begin{equation}
	\ddot{\phi}_0+3H_0\dot{\phi}_0-3\bar{h}H_0\dot{\phi}_0\left[2\ddot{\phi}_0-(q_0-2)H_0\dot{\phi}_0\right]-3\bar{g}\dot{\phi}_0^2\left(\ddot{\phi}_0+H_0\dot{\phi}_0\right)=0, \label{FieldCond}
\end{equation}
with the cosmic deceleration parameter $q$ \cite{Visser:2003vq}. Its best-fit value according to Planck data lies at $q_0\simeq -0.55$ \cite{Planck:2018vyg}, which, however, differs at the 2$\sigma$-level from other recent measurements \cite{Camarena:2019moy}. 
The equations of motion together with the two constraints Eq.~\eqref{wcond} and Eq.~\eqref{Friedcond} allow us to express $\bar{h}$ as a function of $\bar{g}$, $\bar{h}(\bar{g})$, i.e., there is in general a line in the two-dimensional parameter space of couplings, on which the model can describe dynamical dark energy.\\

In addition, there is one more phenomenological constraint on the couplings, which is expressed in terms of the braiding parameter. This is one of the four Horndeski parameters \cite{Bellini:2014fua}.
These are the kineticity $\alpha_K$, braiding $\alpha_B$, Planck-mass run rate $\alpha_M$, and tensor speed excess $\alpha_T$. For our kinetic braiding model \eqref{effac}, we started from the assumption that $\alpha_T=0$. Additionally, $\alpha_M$ vanishes after imposing $G_{4,\chi}=0=G_{4,\phi}$. The other two parameters are given by \cite{Bellini:2014fua}
\begin{align}
		\alpha_K=\,&\frac{8\pi\bar{G}}{H^2 
		}\left[2\chi\left(G_{2,\chi}+2\chi G_{2,\chi\chi}-2G_{3,\phi}-2\chi G_{3,\phi\chi}\right)+12\dot{\phi}\chi H\left(G_{3,\chi}+\chi G_{3,\chi\chi}\right)\right],\label{alphak}\\
		\alpha_B=\,&\frac{ 16\pi\bar{G}\dot{\phi}}{H
		}\left(\chi G_{3,\chi}-G_{4,\phi}\right).\label{alphab}
\end{align}
We focus on $\alpha_B$, because $\alpha_K$ is largely unconstrained \cite{Baker:2020apq}. The braiding parameter is bounded by observational data from cosmological surveys to $|\alpha_B|<1$, see \cite{Noller:2018wyv,Frusciante:2019xia,Noller:2020afd,Traykova:2021hbr}. In addition, the constraint $|\alpha_B + \alpha_M|<10^{-2}$, which arises by demanding stability of perturbations \cite{Creminelli:2019kjy}, translates into 
\be
|\alpha_B|<10^{-2},\label{eq:stabconstraint}
\ee
 in our case, where $\alpha_M=0$.

Translating $G_{3}$ and $G_4$ into the couplings in our setup and assuming vanishing spatial gradients of $\phi$, such that only its time derivative $\dot{\phi}$ remains, we obtain
\begin{align}
		\alpha_B=\,&8\pi\bar{G}\bar{h}\frac{\dot{\phi}^3}{H
		}. \label{braidingparameter}
\end{align}

We can now combine the conditions Eq.~\eqref{wcond} and Eq.~\eqref{Friedcond} that arise from the dark-energy constraint with the constraint on the braiding parameter Eq.~\eqref{eq:stabconstraint} and the equations of motion Eq.~\eqref{FieldCond} to obtain constraints on the couplings.

We first consider the two asymptotic regimes of large and small $\bar{g}$.
The regime $\bar{g} \rightarrow 0$ has previously been analyzed in \cite{Deffayet:2010qz}.
For this case, Eq.~\eqref{wcond} becomes
\be
\ddot{\phi}_0= \frac{w_0-1}{2\bar{h}} -3 w_0H_0 \dot{\phi}_0.
\ee
In turn, this can be substituted into the equations of motion, 
\bea
\frac{w_0-1}{2\bar{h}} +6(1-w_0) H_0 \dot{\phi}_0
+3\bar{h}H^2_0\dot{\phi}_0^2(6w_0+q_0-2)=0,
\eea
which are now of second order in $\dot{\phi}_0$, such that there are two solutions for $\dot{\phi}_0$. These two solutions are
\be
\dot{\phi}_{0,\, 1/2}=\left[6H_0\bar{h}\left( 1\pm\frac{\sqrt{4+q_0}}{\sqrt{6}\sqrt{1-w_0}}\right)\right]^{-1}.\label{phidotofhg0}
\ee
Plugging the solutions into Eq.~\eqref{Friedcond}, we obtain the value of $\bar{h}\simeq \pm 1.7\times 10^{35}{\rm eV^{-3}}\sim\pm\sqrt{\bar{G}}/H_0^2$, thus providing the exact value for an estimate given in \cite{Deffayet:2010qz}.

The above determines a value of $\bar{h}$ for which the braiding parameter is consistent with observations. However, $\bar{h}$ is actually determined independently by the full set of equations specifying the system.
We can use Eq.~\eqref{Friedcond} in addition to Eq.~\eqref{phidotofhg0} to obtain a predicted value for the braiding parameter.
Thus, for both solutions the absolute value of the braiding parameter at present takes the value
\begin{equation}\label{alphabsmallg}
 	|\alpha_{B,0}|=\sqrt{\frac{6(1-w_0)}{4+q_0}}\Omega_{\Lambda,0}\simeq 1.3.
\end{equation}
By virtue of the constraint in Eq.~\eqref{eq:stabconstraint}, this rules out the model in the limit $\bar{g}\rightarrow 0$, see also \cite{Traykova:2021hbr}.

Next, we consider the opposite regime $\bar{g}\rightarrow \infty$ for which the kinetic term in the action \eqref{LorAc} is negligible. From Eq.~\eqref{wcond}, we can thus solve for $\ddot{\phi}$ as
\begin{equation}
	\ddot{\phi}=\frac{1-3w_0}{4}\frac{\bar{g}}{\bar{h}}\dot{\phi}^2-3H_0w_0\dot{\phi},
	\label{wcondglarge}
\end{equation}
where $\bar{h}\neq 0$ was assumed. Substituting this result into the equation of motion Eq.~\eqref{FieldCond}, we obtain
\begin{align}
	\dot{\phi}_0^2&\left[ \dot{\phi}_0^2 \frac{3\bar{g}^2}{4\bar{h}} \left(3 w_0-1\right)+\frac{9}{2} \dot{\phi}_0 \bar{g} H_0 \left(3 w_0-1\right)+3\bar{h}H_0(6w_0+q_0-2)\right]=0,
\end{align}
which can be solved for $\dot{\phi}_0.$ Two of the solutions of this quartic equation are $\dot{\phi}_0=0,$ which implies vanishing dark energy, see Eq.~\eqref{Friedcond}. The other two solutions are
\begin{equation}
	\dot{\phi}_{0,1/2}=\frac{\bar{h}H_0}{\bar{g}}\left(-3\pm{\rm sgn}(\bar{g})\sqrt{\frac{1+4q_0-3w_0}{1-3w_0}}\right)\label{phidotsollargeg},
\end{equation}
with the sign-function ${\rm sgn}(x).$ These solutions are only real if $q_0\gtrsim -1$ which is satisfied by the best-fit result of Planck 2018 data \cite{Planck:2018vyg} but not by independent determinations of it \cite{Camarena:2019moy}. 

Plugging this result into the large-$g$ variant of Eq.~\eqref{Friedcond}
\begin{equation}
	\bar{h}=
	-\frac{H_0\Omega_{\Lambda,0}}{32\pi 
	 \bar{G}\dot{\phi}^3}-\frac{\bar{g}}{4H_0}\dot{\phi},\label{hofphid}
\end{equation}
and considering the dependence $\dot{\phi}\propto \bar{h}$ in Eq.~\eqref{phidotsollargeg}, we obtain two degenerate equations of fourth order in $\bar{h}$. Due to their complexity, we will not display them here. 
It suffices to say that each equation has two solutions for $\bar{h}$ with opposite signs. Those are real \emph{only if} $\bar{g}>0$ and scale as $\bar{h}\sim \pm\bar{g}^{3/4}.$ Thus, the cubic coupling is large in the large-$\bar{g}$ regime -- heuristically, this has to be the case, for the cubic term has to compensate for the large contribution of the quartic term to the energy density.

All of the four solutions for the cubic coupling imply a constant braiding parameter as a function of $\bar{g}$ in the large-$\bar{g}$ regime. Further, Eq.~\eqref{hofphid} implies $\alpha_B\sim\bar{h}^4$, such that the value of the parameter is independent of the sign of $\bar{h}.$ Thus, the pairs of solutions for $\bar{h}$ translate to only two solutions for $\alpha_{B,0}$ 
\begin{equation}\label{alphablargeg}
	\alpha_{B,0}=\frac{\Omega_{\Lambda,0}}{q_0}\left(1-3w_0\pm\sqrt{(1-3w_0)(1+4q_0-3w_0)}\right)\simeq\begin{cases}
		&- 1.7\\
		&- 8.4.
	\end{cases}
\end{equation}
Thus, by Eq.~\eqref{eq:stabconstraint}, not only is the theory ruled out in the limit $\bar{g}\rightarrow 0,$ but also in the large-$\bar{g}$ regime. The only remaining possibility is intermediate $\bar{g}$, which we explore now.

The braiding parameter is negative at large $\bar{g}$ (\cf Eq.~\eqref{alphablargeg}), while it is positive for vanishing $\bar{g}$ (\cf Eq.~\eqref{alphabsmallg}). Thus, if the result is continuous, it must change sign in between. In this regime, $\alpha_B$ may be sufficiently small, leaving open a window within which the stability constraint \eqref{eq:stabconstraint} may be satisfied. Since only positive values of $\bar{g}$ lead to real solutions at large coupling, this reasoning is only consistent for those.

As an aside, it is curious that there are four solutions for $\dot{\phi}_0$ in the large-$\bar{g}$ limit, while there are only two in
 the small-coupling regime. To understand the reason and to find the point where the braiding parameter changes sign, we will now analyze the intermediate region, where the kinetic term in the action is of the same order as the cubic and the quartic interaction terms.

\begin{figure}[htb!]
	\includegraphics[width=\linewidth]{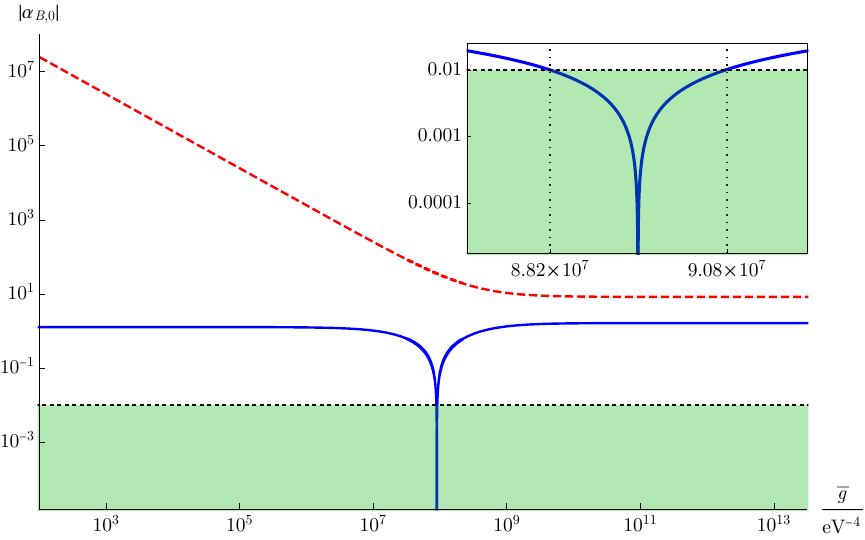}
	\caption{We show the absolute value of the braiding parameter $\alpha_B$ as a function of the quartic coupling $\bar{g}$ (for $\bar{g}>0$) for the two real pairs of solutions of the system of Eqs.~\eqref{wcond}, \eqref{Friedcond} and \eqref{FieldCond}. The cosmological parameters $q_0,$ $H_0,$ $\omega_0$ and $\Omega_{\Lambda,0}$ are set to the best-fit values of the Planck 2018 results \cite{Planck:2018vyg}. The solutions are depicted by the dashed red and solid blue lines, while the dotted black line designates the stability constraint \eqref{eq:stabconstraint} which is satisfied in the green region. The inset plot zooms in on coupling values at which the braiding parameter changes sign for the blue solution. Within the inset plot, the values of the coupling at which the stability constraint is saturated are indicated by vertical, dotted, black lines. \label{Fig:alphaB}}
\end{figure}

\begin{figure}[htb!]
	\includegraphics[width=\linewidth]{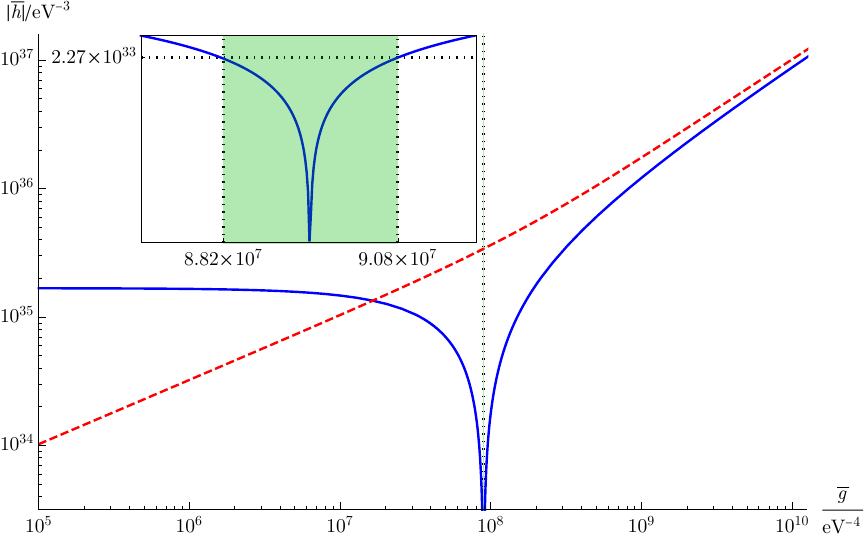}
	\caption{We show the absolute value of the dimensionless cubic coupling at present $|\bar{h}|$ as a function of the quartic coupling $\bar{g}$ (for $\bar{g}>0$) for the two real pairs of solutions of the system of Eqs.~\eqref{wcond}, \eqref{Friedcond} and \eqref{FieldCond}. The cosmological parameters $q_0,$ $H_0,$ $\omega_0$ and $\Omega_{\Lambda,0}$ are set to the best-fit values of the Planck 2018 results \cite{Planck:2018vyg}. Using the same colour code as in Fig.~\ref{Fig:alphaB}, the two pairs of solutions are depicted by red dashed and blue solid lines, while the green region symbolizes the values of the quartic coupling at which the blue solution satisfies the stability bound \eqref{eq:stabconstraint}. The inset plot zooms in on this region of interest. Within the inset plot, the values of the quartic coupling at which the stability constraint is saturated are indicated by vertical, dotted, black lines. This translates as a bound on the cubic coupling delineated by the horizontal, black, dotted line. \label{Fig:hg}}
\end{figure}

\begin{figure}[htb!]
	\includegraphics[width=\linewidth]{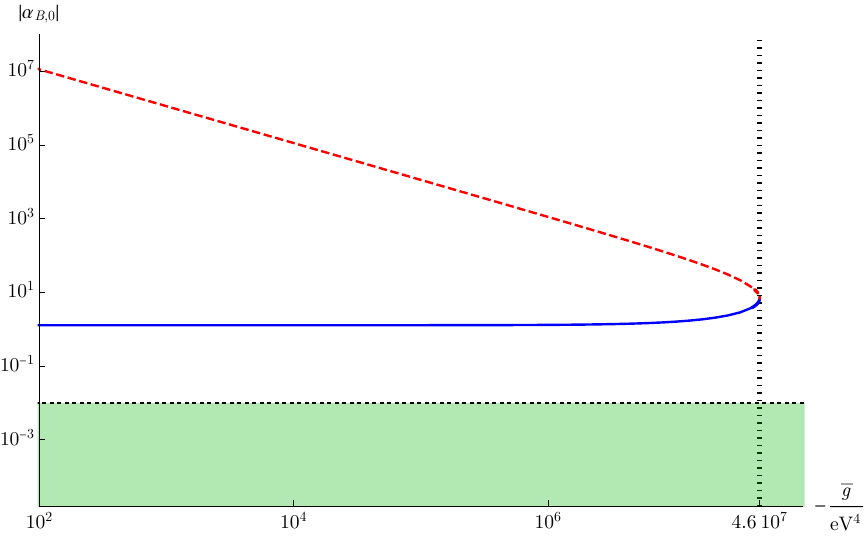}
	\caption{We show the braiding parameter $\alpha_B$ as a function of the quartic coupling $\bar{g}$ (for $\bar{g}<0$) for the two real pairs of solutions of the system of Eqs.~\eqref{wcond}, \eqref{Friedcond} and \eqref{FieldCond}. The cosmological parameters $q_0,$ $H_0,$ $\omega_0$ and $\Omega_{\Lambda,0}$ are approximated by the best-fit values of the Planck 2018 results \cite{Planck:2018vyg}. The solutions are depicted by the dashed red and solid blue lines, while the dotted black line designates the stability constraint \eqref{eq:stabconstraint} which is satisfied in the green region. \label{Fig:alphaBnegg}}
\end{figure}

Even though the nonlinear system made up of Eqs.~\eqref{wcond}, \eqref{Friedcond} and \eqref{FieldCond} can be  solved analytically for $\bar{h},$ $\dot{\phi}$ and $\ddot{\phi},$ an explicit display of the solutions, due to their complexity, hardly leads to a better understanding of their behavior. Therefore, we only plot the resulting graphs for $|\alpha_B|$ and $|\bar{h}|$ as functions of $\bar{g}>0$ in Figs.~\ref{Fig:alphaB} and \ref{Fig:hg}, respectively, and $|\alpha_B|$ as a function of $\bar{g}<0$ in Fig.~\ref{Fig:alphaBnegg}.  

First, we compare some properties of the general solution to the ones obtained in the asymptotic regimes:
\begin{itemize}
	\item There are four real solutions for $\dot{\phi}_0$, \ie two pairs, each with opposite signs for $h$ but equal signs for $\alpha_B.$ However, one pair of solutions, indicated by the red, dashed line in both figures becomes unphysical at vanishing quartic coupling: Even though the cubic and the quartic coupling vanish (see Fig.~\ref{Fig:hg}), the braiding parameter diverges (see Fig.~\ref{Fig:alphaB}). This is why we only found two solutions in that regime. 
	\item The braiding parameter is in fact independent of $\bar{g}$ in the limits $\bar{g}\rightarrow 0$ and $\bar{g}\rightarrow \infty,$ where those exist (\cf Fig.~\ref{Fig:alphaB}), as indicated by the analysis of the asymptotic behaviours.
	\item The cubic coupling indeed scales as $|h|\sim |g|^{3/4}$  for all solutions (\cf Fig.~\ref{Fig:hg}) at large quartic couplings.
	\item There are no real solutions for negative $\bar{g}$ (\cf Fig.~\ref{Fig:alphaBnegg}) beyond the characteristic scale of the problem ($\bar{g}\sim 10^7{\rm eV}^{-4}$), \ie in the limit $|\bar{g}|\rightarrow \infty$, corroborating the asymptotic result.
\end{itemize}

Next, we consider the intermediate region of quartic couplings $\bar{g}\in [10^7{\rm eV^{-4}},10^9{\rm eV^{-4}}].$ As can be gathered from Fig.~\ref{Fig:alphaB}, for the pair of solutions with well-defined small-$\bar{g}$ limit, displayed in blue, the braiding parameter indeed changes sign in this interval for the coupling. Thus, the stability constraint \eqref{eq:stabconstraint} is satisfied for a narrow range of values of $\bar{g}$ zoomed in on in the inset plot of Fig.~\ref{Fig:alphaB}. While the braiding parameter indeed changes sign for positive quartic couplings, an inspection of Fig.~\ref{Fig:alphaBnegg} shows, that it does not for the negative ones, which had been expected from the asymptotic analysis. Summarizing, we obtain a tight bound on $\bar{g}$
\begin{equation}
	\bar{g}\simeq 9\times 10^{7}\text{eV}^{-4}\simeq \frac{1}{H_0^2m_p^2}.\label{gConst}
\end{equation}

Moreover, the cubic coupling changes sign in exactly the region which is allowed by the stability constraint \eqref{eq:stabconstraint} (see Fig.~\ref{Fig:hg}). Additionally, recall that a vanishing cubic coupling leads to a negative cosmological constant (\cf Eq. \eqref{neccosmconst}). Therefore, it is bounded as
\begin{equation}
	0<|\bar{h}|\lesssim 2.3\times 10^{33}\text{eV}^{-3}\simeq 10^{-4}\frac{1}{H_0^2m_p}.\label{hConst}
\end{equation}

We will refer back to the constraints \eqref{gConst} and \eqref{hConst} in the discussion after we present the results for the chosen truncation in the asymptotically safe framework.

\section{Horndeski-like model and quantum gravity in the asymptotic-safety paradigm and effective-field theory}
\label{sec:resultsAS}
 
\subsection{Beta functions}
In this section, we display the beta functions associated with the chosen truncation. We further search for fixed points and analyze their critical exponents, to test whether the asymptotic-safety paradigm can accommodate this specific Horndeski-like model. The choices of regulator and gauge can be found in App.~\ref{app:beta}. 

The RG-flow for the couplings $h$ and $g$ is described by the following beta functions
	\begin{align}
		\beta_{h}=\,& 3 h + \dfrac{g}{4\pi^2}h + G\left[\frac{7 }{4 \pi  (3-4 \Lambda )}+\frac{9 }{2 \pi  (3-4 \Lambda )^2}-\frac{5 }{9 \pi  (1-2 \Lambda )}-\frac{20 }{9 \pi  (1-2 \Lambda )^2}\right]h, \label{betah}\\
		\beta_{g}=\,& 4 g  + \dfrac{9}{32\pi^2}g^2  + \dfrac{G}{6\pi}\left[\frac{8 }{ (3-4 \Lambda )}+\frac{15 }{ (3-4 \Lambda )^2}-\frac{20 }{(1-2 \Lambda )^2}\right]g \nn \\
		& +\dfrac{5  h^2 G}{24\pi}\left[\frac{1 }{ (1-2 \Lambda )} + \frac{1 }{  (1-2 \Lambda )^2}-\frac{3}{ (3-4 \Lambda )}-\frac{9}{ (3-4 \Lambda )^2}\right] \nn \\
		 & + \frac{320 G^2}{9 (1-2 \Lambda )^3} +\frac{2 G^2}{(3-4 \Lambda )^2}+\frac{6 G^2}{(3-4 \Lambda )^3}.\label{betag}
	\end{align}
The Newton coupling $G$ and cosmological constant $\Lambda$ enter these beta functions. Supplementing the system with $\beta_G$ and $\beta_{\Lambda}$, we can search for fixed-point solutions of the full truncation. Alternatively, we can leave $G$ and $\Lambda$ as free parameters. This provides additional insight, because it allows us to test whether different fixed-point values for $G$ and $\Lambda$ could lead to different results for $h$ and $g$. Therefore, we mostly follow this route in the present paper.

In Eq.~\eqref{betah} and \eqref{betag}, we neglect the gravitational anomalous dimension as well as the scalar anomalous dimension arising from regulator insertions, which holds as long as the said anomalous dimensions are small. As a result, the beta functions depend only polynomially on $h$ and $g,$ making at least part of the analysis amenable to analytic treatments.

To obtain the beta functions, we have used self-written Mathematica codes based on the packages \textit{xAct}  \cite{Brizuela:2008ra,Martin-Garcia:2007bqa,MartinGarcia:2008qz}, \textit{FormTracer} \cite{Cyrol:2016zqb}, \textit{DoFun} \cite{Huber:2019dkb} and \textit{Package-X} \cite{Patel:2016fam}. For additional details on the beta functions (and anomalous dimensions) see App.~\ref{app:beta}.

\subsection{Fixed-point candidates and asymptotic safety}

In the following, we keep the fixed-point values of the Newton coupling, $G_{\ast}$, and cosmological constant, $\Lambda_{\ast}$ as free parameters, i.e., do not calculate them from the corresponding beta functions.

The beta functions \eqref{betah} and \eqref{betag} have up to four distinct zeros. However, a zero of a beta function does not automatically correspond to a fixed point, because spurious zeros exist in truncations, as we discuss below.

\subsubsection{Fixed-point candidates in the absence of gravity}\label{sec:FPswoGrav}
As a first step, we consider the case without gravity, i.e., $G = 0$, for which the system of beta functions has two zeros, one at $h_{\ast}=0=g_{\ast}$, i.e., the Gaussian fixed point, and one at $h_{\ast}=0, g_{\ast}=- \frac{128\pi^2}{9}$. For the latter, the critical exponents are $\theta_1=4$, $\theta_2=5/9$. This is a very significant shift compared to the critical exponents at the Gaussian fixed point $\theta_1=-4$, $\theta_2=-3$. Such a large shift implies that the assumption of a near-perturbative fixed point, at which the canonical critical exponent provides a robust estimate of whether a coupling is relevant, breaks down. Therefore, our truncation is not well-suited to characterize the interacting zero of the beta functions, or, indeed, to determine whether this is an actual fixed point.

\subsubsection{Fixed-point candidates in the presence of gravity at $\Lambda_{\ast}=0$}\label{sec:lambdazero}
Next, we turn on gravitational fluctuations, but focus on the special case $\Lambda_{\ast}=0$. Phenomenologically, the case of vanishing IR-value of the cosmological constant is of particular interest for Horndeski theories. A vanishing IR value is, however, not connected to a vanishing UV value.  Therefore, the case $\Lambda_{\ast}=0$, where the UV value is set to zero, is not motivated by phenomenology, but instead of interest due to a special algebraic property of the beta functions. In this case, the contribution $\sim h^2 G$ in Eq.~\eqref{betag} is absent. Thus, the system of beta functions remains quadratic in $g$ and linear in $h$,
\begin{align}
	\beta_h=&\,3h+\frac{g}{4\pi^2}h-\frac{61G}{36\pi}h,\label{betahnol}\\
	\beta_g=&\,4g+\frac{9}{32\pi^2}g^2-\frac{47G}{18\pi}g+36 G^2.\label{betagnol}
\end{align}
The beta function for $g$ is independent of $h$; thus we first investigate the implications of $\beta_g$.
It still exhibits two zeros. Both are shifted compared to their vanishing-gravity counterparts in Sec.~\ref{sec:FPswoGrav}.

Depending on the value of $G_*$, there are two, one or no real solutions for $g_*$, reflecting the \textit{weak-gravity bound} as discussed in \cite{deBrito:2021pyi}:
\be 
g_{\ast\,\pm}=\frac{8}{81} \pi  \left(47 G_*-72 \pi \pm \sqrt{-10913 G_*^2-6768 \pi  G_*+ (72 \pi )^2}\right).
\ee
 Gravitational fluctuations shift the Gaussian fixed point towards negative values, and as $G_*$ increases the two fixed-point candidates approach each other. For a critical value $G_*=G_{\ast,\, \text{crit}}$, the two fixed-point candidates collide such that there is only one real solution for $g_*$. At even larger $G_*$, there is no fixed-point solution\footnote{The same mechanism holds for non-vanishing values of $\Lambda_*$, as can be seen in Fig.~\ref{fig:FixedPointContour}.}. Consequently, the weak-gravity bound constrains the available parameter space for $G_{\ast}$ to the regime of sufficiently weakly-interacting gravity. The critical value of $G_{\ast\, \rm crit}$, beyond which gravity leaves the weakly-coupled (in the sense of the weak-gravity bound) regime, is
 \be
 G_{\ast\, \rm crit}= \frac{72\pi}{10913} \left(81 \sqrt{2}- 47 \right) \approx 1.4\,.
 \ee
Next, we add $\beta_h$ to the system. The result
\be
h_{\ast}=0
\ee
holds for all values of $G_{\ast}$. This is rooted in symmetry-considerations: the coupling $h$ breaks the $\mathbb{Z}_2$ symmetry of the kinetic term and the other interaction; thus, fluctuations do not generate it in the fixed-point regime and its beta function is proportional to $h$, as is always the case with symmetry-protected interactions.

A vanishing fixed-point value of a coupling does not imply that its IR value must vanish; indeed, asymptotically free non-Abelian gauge couplings are a classical example of a vanishing fixed-point value and non-vanishing IR value. The prerequisite for this is a positive critical exponent (or, if the critical exponent is zero, the leading-order term in the beta function must come with a negative sign). To connect the fixed-point candidates to IR phenomenology, we thus focus on the critical exponent related to $h$. Since $\beta_g$ does not depend on $h$, the stability matrix is upper/lower triangular, and thus 
\bea
\theta_{0,h} &=& - \frac{\partial \beta_h}{\partial h}\Big|_{h= h_{\ast}=0, \, g=g_{\ast\, +}}\nonumber\\
&=& -3 + \frac{61}{36 \pi} G_{\ast}- \frac{84 G - 144 \pi + 2\sqrt{-10913 G_{\ast}^2-6768\pi\, G_{\ast}+ 5184 \pi^2}}{81\pi}.
\eea
As the fixed-point candidate at large negative values of $g_{\ast}$ is not reliably characterizable within our truncation, we limit ourselves to discussing the shifted Gaussian fixed point, \ie, $g_{\ast}=g_{\ast\,+}$. For $G_{\ast} \rightarrow 0$, $g_{\ast\, +}\rightarrow 0$, \ie, gravitational fluctuations shift the Gaussian fixed point to an interacting fixed point. At this fixed-point value for $g$, the critical exponent is monotonically increasing as a function of $G_{\ast}$, from $\theta_{0,h}=-3$ at $G_{\ast}=0$ to $\theta_{0,h} \approx -1$ at $G_{\ast}=G_{\ast\, \rm crit}$.

\begin{figure}[!t]
	\centering
	\includegraphics[]{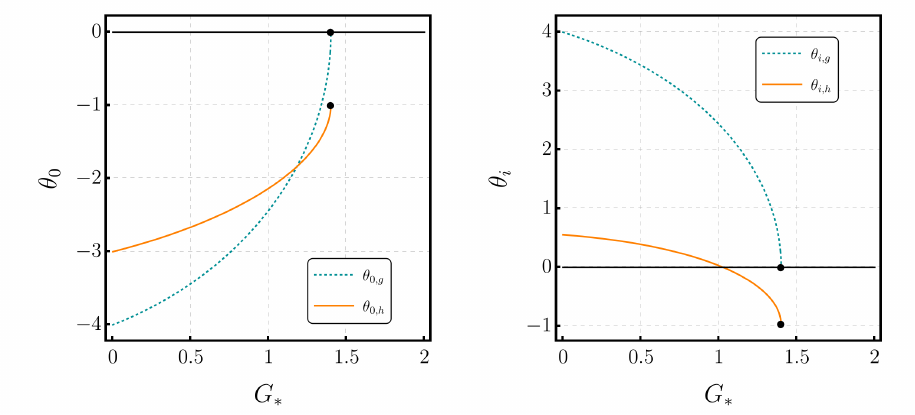}
	\caption{The critical exponents at the shifted Gaussian fixed point (left panel). $\theta_{0,g}=0$ at $G_{\ast}=G_{\ast\, \rm crit}$, because the shifted Gaussian fixed point is degenerate with the interacting fixed-point candidate for this value; the critical exponents of which are shown in the right panel. The dots at the end of the trajectories indicate the weak-gravity bound.} 
	\label{fig:CritExps}
\end{figure}

The gravitational effect therefore goes in the desired direction, i.e., it renders $h$ less irrelevant than indicated by canonical scaling. However, a change in $\theta_{0,h}$ large enough to overcome the canonical scaling, such that $\theta_{0,h}>0$, is not achieved before $G_{\ast}$ reaches the weak-gravity bound, cf.~Fig.~\ref{fig:CritExps}. As a consequence, $h=0$ holds not only in the fixed-point regime, but at all scales. 

In turn, as commented in Sec.~\ref{sec:obs}, $h=0$ implies that the given kinetic braiding model cannot describe dark energy, see Eq.~\eqref{alphab}.\\

In summary, at the shifted Gaussian fixed point, asymptotic safety predicts a braiding parameter that is \emph{incompatible} with the stability bound \eqref{eq:stabconstraint}. This prediction is subject to a number of caveats, most importantly including our choice of truncation. In order to explore whether operators that we have neglected here could produce a large enough shift in $\theta_{0,h}$ to make $h$ relevant before $G_{\ast} = G_{\ast\, \rm crit}$ is reached, we perform a parametric study of the effect of additional operators in App.~\ref{app:higherorder}. However, these considerations do not change the result. A similar conclusion was found in \cite{Melville:2022ykg}, where positivity bounds were used to constrain low-energy EFTs by imposing unitarity, causality and locality in the UV. It is interesting to speculate whether this similarity of results can be attributed to asymptotic safety satisfying unitarity, causality and locality.

\subsubsection{Fixed-point candidates in the presence of gravity at $\Lambda_{\ast}\neq0$}\label{sec:lambdaneqzero}

\begin{figure}[!t]
	\centering
	\includegraphics[]{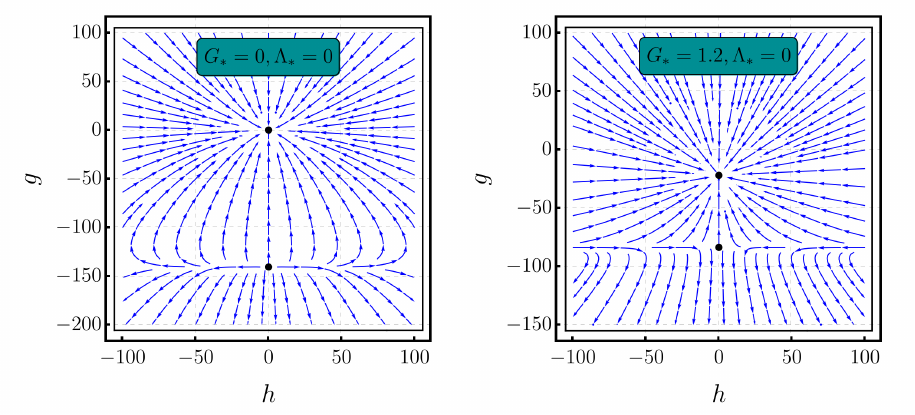}\\
		\includegraphics[]{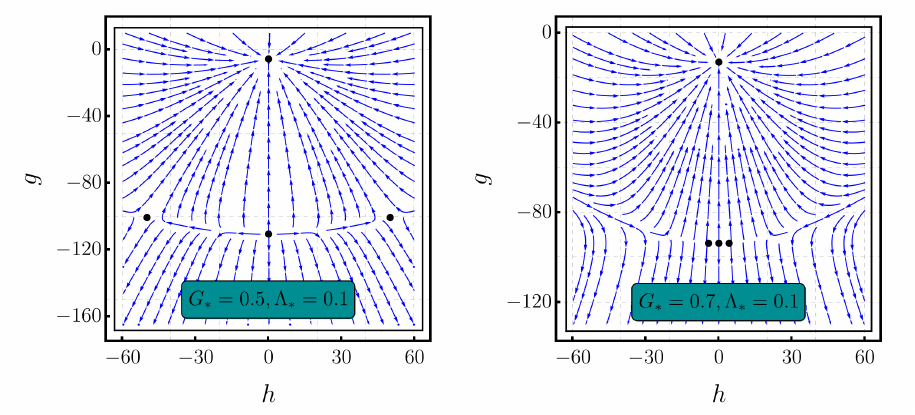}
	\caption{RG flow towards the IR for $\Lambda_*=0$ (upper panels) and $\Lambda_{\ast}=0.1$ (lower panels) and different values of $G$. 
	The (shifted) Gaussian fixed point is the uppermost one in all panels, and fully IR attractive with two irrelevant couplings. Additional fixed-point candidates feature at least one relevant direction and cannot reliably be characterized in our truncation because they do not obey the assumption of near-canonical scaling that our truncation is based on.
	} 
	\label{fig:StreamPlots}
\end{figure}

At $\Lambda_{\ast} \neq 0$, the polynomial equations Eq.~\ref{betah} and \ref{betag} are of fourth order; thus the general case has four fixed-point candidates, which are illustrated in Fig.~\ref{fig:StreamPlots}, where two of the four candidates, ($h_*\neq 0, g_*\neq 0$), disappear for certain values of $\Lambda_*$ including at $\Lambda_*=0$. For the interacting fixed-point candidates with at least one relevant direction, our truncation is too small for a reliable characterization, since they can be artefacts of the approximation; our focus therefore remains on the shifted Gaussian fixed point.

\begin{figure}[]
	\centering
\begin{minipage}{.475\linewidth}
		\centering
	\includegraphics[width=\linewidth]{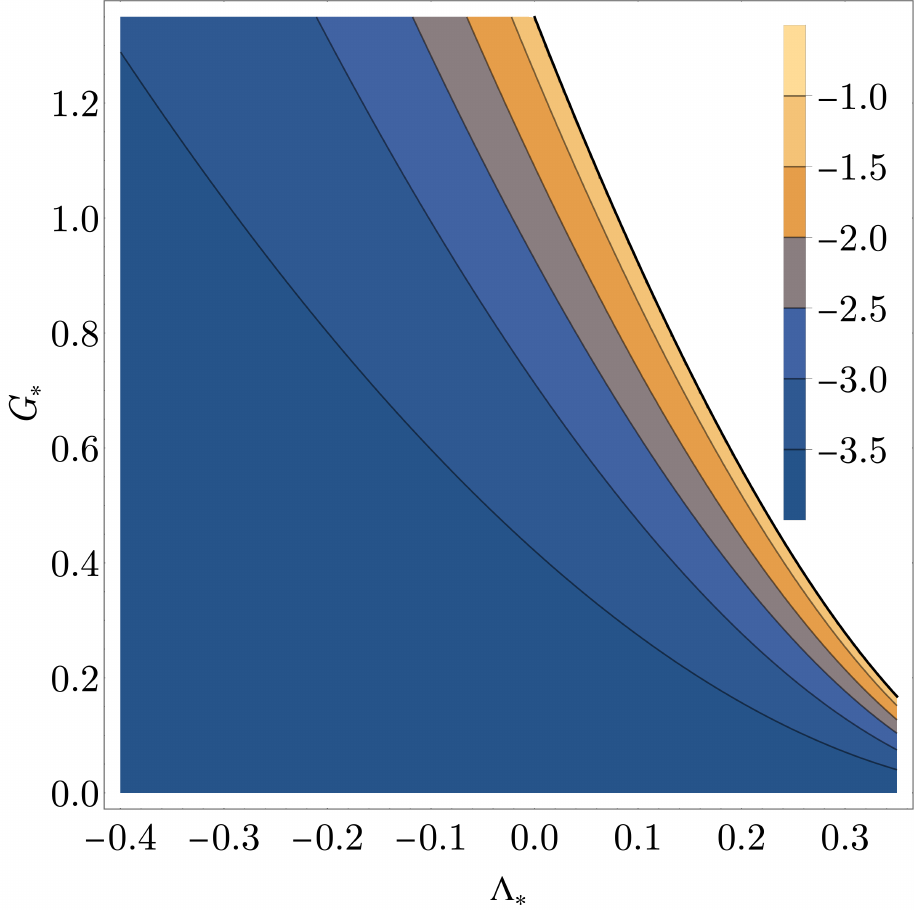}
\end{minipage} \hspace{.03\linewidth} 	
\begin{minipage}{.475\linewidth}
	\centering
	\includegraphics[width=\linewidth]{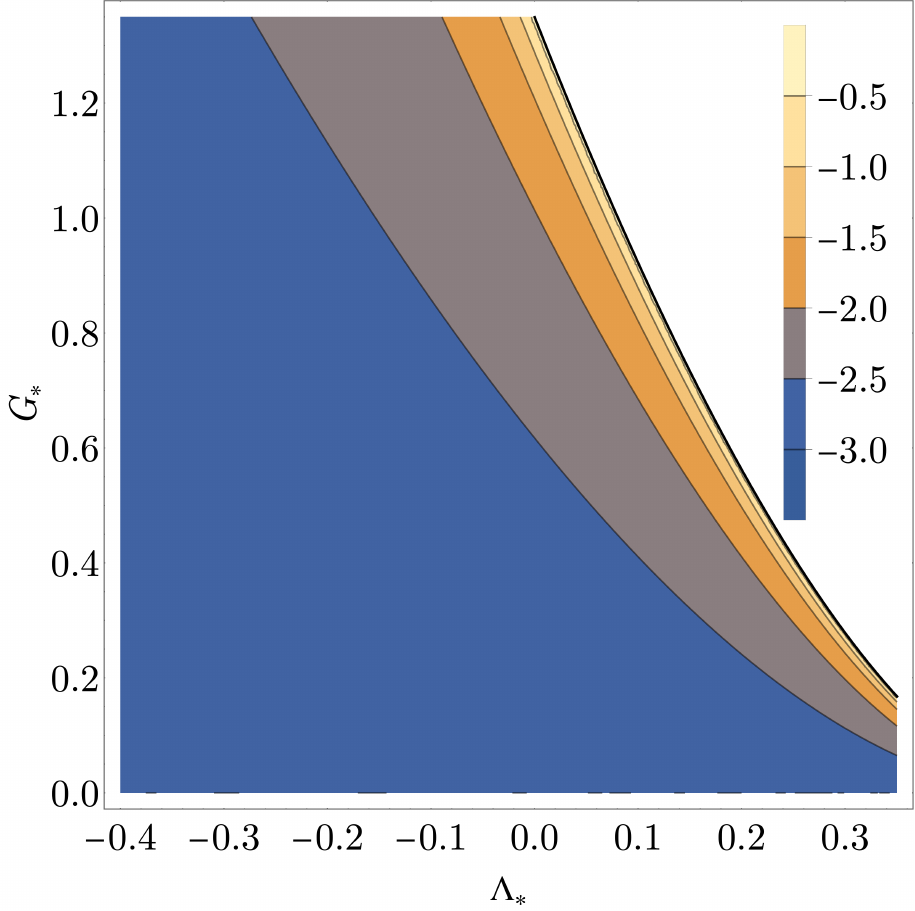}
\end{minipage}
	\caption{Contour plots displaying the two critical exponents $\theta_{0,h}$ (left panel) and $\theta_{0,g}$ (right panel) of the shifted Gaussian fixed point as functions of the gravitational couplings. The white areas are unattainable due to the weak gravity bound at $\theta_{0,h}=-1,$ $\theta_{0,g}=0$ which is indicated by the thick, black line.} 
\label{fig:FixedPointContour}  	
\end{figure}

The situation at $\Lambda_{\ast}=0$ is characteristic for the more general case: increasing $G_{\ast}$ shifts $\theta_{0,h}$ towards zero, i.e., gravitational fluctuations are shifting $h$ towards relevance. However, the weak-gravity bound is encountered before $\theta_{0,h}>0$, \cf~Fig.~\ref{fig:FixedPointContour}. Accordingly, we conclude that $h=0$ in the IR is a prediction that holds irrespective of the gravitational fixed-point values $G_{\ast},$ $\Lambda_{\ast}$. In particular, supplementing our calculation with gravitational beta functions in the presence of Standard-Model matter, as in \cite{Dona:2013qba, Eichhorn:2020sbo}, results in fixed-point values within a regime where the shifted Gaussian fixed point exists and predicts $h=0$ at all scales. 

This prediction is subject to caveats, e.g., taking into account additional gravitational interactions (e.g., higher-curvature terms) increases the available parameter space, and a region might open up for non-zero values of higher-order gravitational couplings where $\theta_{0,h}>0$ before the weak-gravity bound is reached.

\subsection{Effective-field theory perspective}
\label{sec:resultsEFT}
Let us now broaden our view to an effective-field theory perspective on quantum gravity. Here, we have in mind a scenario where a quantum field theory of the metric continues to hold up to some new-physics scale $M_{\rm NP}$ (which may even be higher than the Planck scale) but breaks down beyond. There is therefore a regime in which quantum-gravity fluctuations can be described within the FRG formalism.

Within this general setting, we investigate whether there are generic predictions that arise, independent of the fundamental quantum-gravity theory that holds beyond $M_{\rm NP}$. For this investigation, the concept of UV intervals and IR intervals is important. A UV interval for a coupling is an interval of values within which the coupling is assumed to lie at the UV scale, in this case $M_{\rm NP}$. The RG flow maps this UV interval onto a corresponding IR interval. We are interested in understanding whether a large UV interval, \eg, $h,~g \in [-1,1]$ (motivated by a perturbativity constraint) is mapped to a smaller IR interval, such that a prediction $(\bar{h}_{\text{EFT}},\bar{g}_{\text{EFT}})$ emerges for the dimensionful couplings in the IR. These considerations are closely related to predictivity measures for effective asymptotic safety, see \cite{Held:2020kze}.

To demonstrate the concept, we first focus on the subplanckian regime, where gravity fluctuations are negligible. The beta functions can then be integrated analytically, resulting in
\begin{align}
	h(k)=&\frac{h_i\left(\frac{k}{k_i}\right)^3}{\left[1+\frac{9g_i}{128\pi^2}\left[1-\left(\frac{k}{k_i}\right)^4\right]\right]^{8/9}},
	\label{RGtrah}\\
	g(k)=&\frac{g_i\left(\frac{k}{k_i}\right)^4}{1+\frac{9g_i}{128\pi^2}\left[1-\left(\frac{k}{k_i}\right)^4\right]},\label{RGtrag}
\end{align}
with $h(k_i)= h_i$ and $g(k_i)=g_i$ as initial conditions. At $k<k_i$, the denominators quickly become constant, and the numerators exhibit canonical scaling, $h(k) \sim k^3$ and $g(k)\sim k^4$. A UV interval of extent $h_i \in [-1,1]$ is therefore mapped to an IR interval of $h(k)\in[- \tfrac{k^3}{k_i^3},\tfrac{k^3}{k_i^3}]$ and similarly for $g$.

The IR interval can in turn be compared to the observational constraints on the braiding parameters. For this, we
make the transition to dimensionful couplings. In this transition, the denominators in Eqs.~\eqref{RGtrah} and \eqref{RGtrag} are again negligible, such that
\bea
\bar{h}(k) &=& \frac{h_i}{k_i^3},\\
\bar{g}(k)&=& \frac{g_i}{k_i^4},
\eea
i.e., the dimensionful couplings are essentially constant. From this it follows that dimensionless coupling values with absolute values smaller than $\mathcal{O}(1)$ at the Planck scale imply that $|\bar{h}_{\text{EFT}}|<m_p^{-3}$ and $|\bar{g}_{\text{EFT}}|<m_p^{-4}.$ Comparison with the bounds \eqref{gConst} and \eqref{hConst} shows that the EFT-prediction does not challenge the upper bound on $|h_0|,$ undercutting it by a factor $10^{-4}H^2/m_p^2\sim 10^{-57}.$ The quartic coupling, however, results far, \ie by the factor $H^2/m_p^2$, too small to comply with Eq.~\eqref{gConst} and therefore insufficient to provide a dynamical explanation of dark energy. This may be interpreted as an indication that new physics must be present that alters the scaling, in order to make the model phenomenologically viable.

Interestingly, we may also use Eqs.~\eqref{RGtrah} and \eqref{RGtrag} to estimate a scale of new physics for Horndeski theory. This is because of the divergence in both $h(k)$ and $g(k)$ that arises due to the possible zero in the denominator for positive $g_i$. We obtain
\be
\frac{k^2}{k_i^2} = \frac{1}{3}\sqrt{\frac{9 g_i+ 128 \pi^2}{g_i}}.
\ee
For $g_i \sim \mathcal{O}(1)$, the strong-coupling regime sets in essentially at $k_i$. Correspondingly, to extend the theory to significantly higher than $k_i$, $g_i$ has to decrease by four orders of magnitude for each order of magnitude one wishes to gain. Thus, Horndeski-like models which remain weakly coupled up to the Planck scale, where gravity changes the RG flow, are constrained to tiny values of $g$ in the IR, which can hardly be reconciled with the constraint \eqref{gConst}.\\

Next, we include the effect of gravity. We set $\Lambda=0$ and assume $G \approx \rm const$. The latter is motivated by the idea of \emph{effective asymptotic safety} \cite{Percacci:2010af,deAlwis:2019aud,Held:2020kze}, i.e., a regime in which the gravitational fixed point is nearly realized, but not quite. This is a prerequisite for the extension of the EFT regime beyond the Planck scale. Without such an assumption, the scaling $G \sim k^2$ drives gravity into a strongly-coupled regime at or slightly above the Planck scale.\\
Because we work within an EFT setting, there are two regimes of interest, only one of which was available in the asymptotically safe setting: the first is the regime where both couplings are irrelevant in the vicinity of the free fixed point, the second is the regime where both couplings are relevant in the vicinity of the free fixed point.  In the regime in which all couplings are small, it holds that
\bea
h(k) &=& h_i \left(\frac{k}{k_i}\right)^{3-\frac{61 G}{36\pi}},\label{eq:hflowtransplanckian}\\
g(k) &=& g_i \left(\frac{k}{k_i}\right)^{4-\frac{47 G}{18\pi}}.\label{eq:gflowtransplanckian}
\eea 
Gravitational fluctuations counteract the canonical scaling. Even in the regime where both couplings are irrelevant,  the decrease of the dimensionless couplings towards the IR is slower than with gravity. 
The full scale dependence of the couplings is a combination of two powerlaws: the gravity-dressed scaling $3-\frac{61 G}{36\pi}$ (or $4-\frac{47 G}{18\pi}$, respectively), above the Planck scale, combined with the canonical scaling $r$ (or 4, respectively), below the Planck scale.
In this regime, the dimensionful couplings increase with scaling exponents $\frac{61 G}{36\pi}$ and $\frac{47 G}{18\pi}$ in the transplanckian regime, respectively, before they become constant.

There is a second regime, in which the couplings are not limited to be small, and the terms $\sim G^2$ and $\sim g^2$ in Eq.~\eqref{betag} become important. In this regime, the couplings are relevant and the dimensionless couplings behave as follows: a given UV interval of initial conditions at some transplanckian scale is mapped to a larger interval at the Planck scale, which in turn is mapped to a smaller interval at low energies. Whether the final IR interval is larger or smaller than the UV interval, depends on the value of $G$ and the choice of UV and IR scale. 


Finally, we show that it is not possible to reproduce the constraint on \eqref{gConst} either way: for all values limited by a perturbativity constraint for the coupling $g$ at the transplanckian scale, the order of magnitude of its dimensionful counterpart in the IR stays below $10^{-7} {\rm eV}^{-4}$
. To compensate for this feature, it would be necessary to include even larger, possibly non-perturbative gravitational effects, which are beyond the scope of this EFT analysis. 
Contributions from higher-order operators could in principle help, but they are shown to contribute even less, cf. App.~\ref{app:higherorder}.


\section{Conclusions}
\label{sec:results}
Horndeski theories are widely considered extensions of General Relativity, featuring an additional scalar degree of freedom while avoiding Ostrogradski instabilities. In particular, shift-symmetric kinetic braiding models can describe how dark energy emerges dynamically. These models are typically considered at the phenomenological level, with many analyses focusing on classical formulations. Instead, we take a more fundamental point of view and ask whether these models can be embedded into a fundamental quantum field theory within the asymptotic-safety paradigm.

It is crucial that the self-interactions of the scalar field in Horndeski-like theories are canonically higher-order ones. From an RG point of view one therefore generically expects the corresponding couplings to be tiny at low energies. Therefore, it is interesting to ask not just whether Horndeski-like theories can be embedded into asymptotic safety, but also, whether asymptotic safety makes predictions for the couplings which deviate from the expectation based on canonical scaling.

We find a fixed-point candidate for Horndeski-like models, in principle rendering the model asymptotically safe. However, at this fixed point, one of the Horndeski couplings vanishes and is irrelevant. This results in the prediction that the coupling $h$ vanishes at all scales. In other words, the Horndeski-like model cannot be accommodated in asymptotic safety at nonvanishing values of this coupling. In this case, the particular model is incompatible with a positive energy density. 
The very reason for which Horndeski-models are typically introduced, namely to provide a dynamical explanation of dark energy, therefore is not compatible with the asymptotic-safety paradigm, at least for the simple model we consider here and subject to the technical limitations of our study. The accelerated late-time expansion of the universe may therefore be explained by (i) a cosmological constant (for which a ``naturalness" problem can be avoided in a unimodular formulation of asymptotic safety \cite{Eichhorn:2013xr,Eichhorn:2015bna}) or (ii) a different dynamical dark-energy model, see, e.g., \cite{Wetterich:2022ncl}.

Two further implications follow from this result. First, it adds to the existing evidence that the number and type of effective field theories compatible with an asymptotically safe UV completion is limited. This in turn makes observational tests of asymptotic safety possible. Second, it suggests that a UV completion of Horndeski-like, shift-symmetric theories with non-vanishing braiding is not possible within a quantum field theoretic framework, because such a UV completion would require asymptotic safety or freedom, none of which is available.

Going beyond asymptotic safety, we analyze an effective-field theory setting for the quantized Horndeski-like model. We find that for strong enough gravitational fluctuations, the canonically irrelevant self-interactions of the scalar field can become relevant in the quantum-gravity regime, and thus grow in value towards low energies. However, we show that this growth is not large enough and the IR value of the quartic coupling is not compatible with the previously derived constraint, at least for a reasonable range of initial values for the coupling in the UV. As a consequence, Horndeski gravity appears to be incapable of describing dynamical dark energy, under the assumption that there is no subplanckian new physics scale. If there is such a scale, our analysis does not apply, and the required values of the couplings may be realizable. This, however, suggests, that to successfully model dynamical dark energy, Horndeski gravity requires additional degrees of freedom, and, very likely, nonperturbative physics, in order to achieve a significantly different behavior of the couplings than what is expected based on their canonical dimension.

Our conclusions are subject to technical caveats. We work with functional RG techniques, which limits us to exploring Euclidean signature. Additionally, we perform a systematic approximation by neglecting interactions beyond dimension eight, which are generated by the RG flow. In the appendix, we address one of those caveats by including an interaction beyond our original approximation. We find that it does not change our conclusions.

Our study establishes that asymptotic safety has predictive power for Horndeski models. Thus, it will be interesting to explore whether additional couplings from the larger Horndeski family can be predicted within asymptotic safety and whether the predictions are close to being tested observationally. Additionally, it will also be of interest to extend these type of studies beyond the shift-symmetric sector or polynomial truncations. These extensions are left for future investigation.

\acknowledgments
This work is supported by a research grant (29405) from VILLUM FONDEN. We thank Gustavo P.~de Brito for useful discussions. F.~W.~thanks the Quantum Gravity group at CP3-origins, University of Southern Denmark, for the warm hospitality during his stay and acknowledges support by the Polish National Research and Development Center (NCBR) project ''UNIWERSYTET 2.0. --  STREFA KARIERY'', POWR.03.05.00-00-Z064/17-00 (2018-2022) and the COST Action CA18108. R. R. L. dS. thanks Kai Schmitz and the "Particle Cosmology Münster" group at the University of Münster for the hospitality during the very last stages of this work.

\appendix
\section{Methods \label{app:beta}}

This appendix is devoted to the formalism and conventions that let us derive the beta functions \eqref{betah} and \eqref{betag} from the ansatz \eqref{effac}. First, we assume the gravitational background to be parametrized as flat with added graviton corrections to second order in the form
\begin{equation}
g_{\mu\nu}=\delta_{\mu\nu}+\sqrt{32\pi \bar{G}}h_{\mu\nu},
\end{equation}
with the Kronecker delta, \ie Euclidean metric, $\delta_{\mu\nu}.$ Furthermore, we add a gauge fixing term
\begin{equation}
\Gamma_{\text{gf}}=\frac{1}{\alpha}\int\D^4x\sqrt{\det g_{\mu\nu}}g^{\mu\nu}F_\mu F_\nu
\end{equation}
to the effective action, where we introduced the vector-valued function
\begin{equation}
F_\mu=\frac{1}{8\pi G}\left(\nabla^\nu h_{\nu\mu}-\frac{1+\beta}{4}\nabla_\mu h\right).
\end{equation}
In particular, we work in Landau gauge ($\alpha=0$) additionally selecting $\beta=0$ such that the only propagating gravitational mediators are the transverse traceless (spin-2) and trace (spin-0) modes. 

To obtain the effective average action, we further need the regulator terms
\begin{align}
\frac{1}{2}\int\D^4x\sqrt{\det g_{\mu\nu}}h_{\mu\nu}R^{\mu\nu\rho\sigma}_{h,k}h_{\rho\sigma}+\frac{1}{2}\int\D^4x\sqrt{\det g_{\mu\nu}}R_{\phi,k}\phi^2,
\end{align}
with the regulators
\begin{align}
R^{\mu\nu\rho\sigma}_{h,k} (p)=&\left.\frac{\delta^2\Gamma}{\delta h_{\mu\nu}(p)\delta h_{\rho\sigma}(-p)}\right|_{\Phi=0}r(k/p),\\
R_{\phi,k}(p)=&\left.Z_\phi\frac{\delta^2\Gamma}{\delta\phi (p)\delta\phi (-p)}\right|_{\Phi=0}r(k/p).
\end{align}
Here we abbreviated $\Phi=\{h_{\mu\nu},\phi,\Lambda\}$ and introduced the Litim regulator function
\begin{equation}
r(y)\equiv\left(y^2-1\right)\theta \left(y^2-1\right),
\end{equation}
with the Heaviside step function $\theta (x).$

From this starting point, we express the right-hand side of Eq.~\eqref{FRGeq} in terms of the quantities
\begin{align}
\mathcal{P}^{-1}=&\left.\left(\Gamma^{(2)}_k+R_k\right)\right|_{\Phi=0},\\
\mathcal{F}=&\Gamma^{(2)}_k-\left.\Gamma^{(2)}_k\right|_{\Phi=0},
\end{align}
applying the $\mathcal{P}^{-1}\mathcal{F}$-expansion to the flow equation
\begin{equation}
k\partial_k\Gamma_k=\frac{k}{2}\text{Tr}\tilde{\partial}_k\ln\mathcal{P}+\frac{k}{2}\sum_{n=1}^\infty\frac{(-1)^{n+1}}{n}\text{Tr}\tilde{\partial}_k\left(\mathcal{P}_k^{-1}\mathcal{F}\right)^n,
\end{equation}
where the derivative $\tilde{\partial}_k$ only applies to regulators. For finite truncations, the resulting series contributes to finite order. This is to be understood as a nonperturbative vertex expansion at one loop, which can be evaluated diagram by diagram, allowing us to obtain the beta functions. 

\section{Impact of higher order operators}\label{app:higherorder}

As we showed in Sec.~\ref{sec:lambdazero}, assuming the given truncation, the critical exponents around the shifted-Gaussian fixed point are negative for any allowed value of $G_*$ and $\Lambda_*$. Could the inclusion of higher order operators change this result?

\begin{figure}[t]
	\centering
	\includegraphics[]{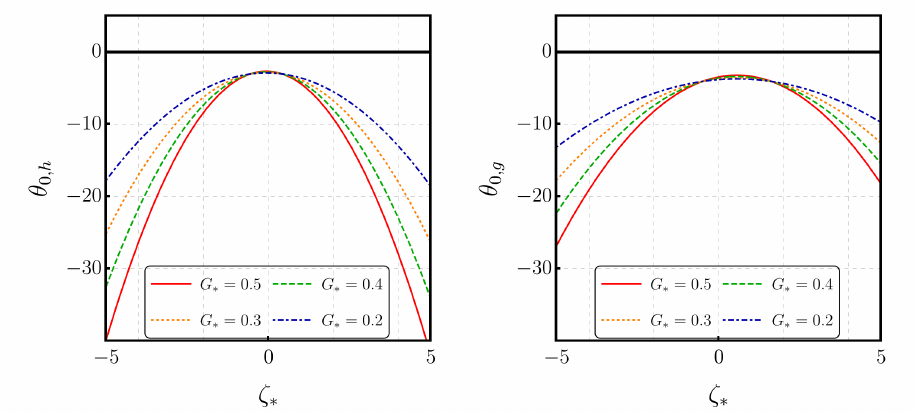}
	\caption{Left (right) panel: critical exponents of the shifted Gaussian fixed point associated with the coupling $h$ ($g$) for different fixed-point values of the gravitational coupling. In all cases, the additional higher-order operator does not change the sign of the critical exponents. }
	\label{fig:CritExps2}
\end{figure}

We expect that contributions from higher operators are negligible because they are canonically irrelevant. For completeness, let us check if quantum gravity fluctuations can change the sign of the critical exponents. In the spirit of Horndeski theories, the next-to-leading-order terms to be considered correspond to the ones in the family $\Lag_4$, \cf Eq. \eqref{L4}. Choosing $G_4(\phi,\chi)_{\text{extra}}=\bar{\zeta} \, \chi$, the corresponding operator reads
\be
\Gamma_{k,\text{extra}}=\int d^4 x\sqrt{\text{det}g_{\mu\nu}} Z_\phi\, \zeta \, k^{-2} \left[\chi\,R + 2\, g^{\mu\nu}g^{\alpha\beta}\left(\nabla_\mu \del_\nu \phi \nabla_\alpha \del_\beta \phi - \nabla_\mu \del_\alpha \phi \nabla_\nu \del_\beta \phi \right)\right],
\ee
which continues to be shift-symmetric and introduces a non-minimal coupling. Since we are interested in the leading-order effect of $\zeta$ in $\beta_h$, we will neglect the back-reaction of the flow of the coupling $\zeta$ on itself and on $\beta_g$. In that vein, the dependence of $\beta_h$ on $\zeta$ is given by
\be
\beta_h\vert_{\zeta}=\dfrac{Gh}{\pi}\left[\frac{5 \zeta  }{12  (1-2 \Lambda )}+\frac{5 \zeta }{12 (1-2 \Lambda )^2}+\frac{161 \zeta }{20  (3-4 \Lambda )}+\frac{339 \zeta }{20   (3-4 \Lambda )^2}+\frac{12 \zeta ^2}{(3-4 \Lambda )}+\frac{18 \zeta ^2}{  (3-4 \Lambda )^2} \right].
\ee
The critical exponent $\theta_{0,h}$ associated with the shifted Gaussian fixed-point $(h_*=0, g_*=g_{+}, \zeta_* )$ is modified but, as the  positive dependence on $\zeta^2$ in $\beta_h$ suggests, the additional terms push the critical exponent to even smaller values. The same is true for $\theta_{0,g}$, \cf Fig. \ref{fig:CritExps2}. Consequently, our result has not changed: the critical exponents are still negative and the directions irrelevant.

\bibliography{references}

\end{document}